# IDENTIFYING HABITABLE EXOPLANETS WITH RADIO TELESCOPES ON THE LUNAR FARSIDE

N. Mahesh[1], J. D. Bowman[2], J. O. Burns[3], S. D. Bale[4], T-C. Chang[6], S. Furlanetto[7], G. Hallinan[1], A. Hegedus[8], J. Mirocha[6], J. Pober[8], R. Polidan[9], D. Rapetti[10], N. Thyagarajan[5], J. Turner[11]. [1]Caltech, Pasadena, CA, [2]Arizona State University, Tempe, AZ, [3]University of Colorado, Boulder, CO, [4]University of California, Berkeley, CA, [5]CSIRO, Perth, W. Australia, [6]JPL/Caltech, Pasadena, CA, 3SSL, [7]UCLA, Los Angeles, CA, [8]Brown University, Providence, RI, [9]Lunar Resources, Houston, TX, [10]USRA, NASA Ames Research Center, Moffett Field, CA,[11]Cornell University, NY.

**1. Science Motivation:** The search for habitable conditions beyond Earth is a top priority in astrophysics. The discovery of habitable exoplanets beyond our solar system will require a suite of instruments providing long-term monitoring for detection (e.g. with space and ground-based radial velocity observations), spectroscopic characterization of atmospheric and surface properties, and eventually deep chronograph-aided observations from e.g. JWST, Roman Space Telescope, and the Habitable Worlds Observatory (HWO). Detection of exoplanet magnetospheres is necessary to identify the most promising targets for follow-up characterization of biosignatures with these assets, and to provide an ensemble of objects for studies of magnetospheric conditions and atmospheric composition. **Strong planetary magnetospheres are critical for retaining atmospheres needed for life**. Enhanced radiative output during stellar flares has been shown to expand and expel atmospheres. Fast, dense winds, especially from young stars and low-mass M-dwarf stars (the most common stellar type), are capable of compressing magnetospheres and exposing atmospheres to mass loss (Kodachenko et al. 2007, Lammer et al. 2007), particularly during coronal mass ejections (CMEs). This has been tested in our Solar System. MAVEN found evidence that ion loss from solar winds strongly depleted the early Martian atmosphere (Jakovsky et al. 2015).

**Only observations of low-frequency radio emission will distinguish exoplanet magnetospheres** (Hallinan et al. 2021). Exoplanet radio bursts indicate the presence and strength of a magnetic field. For example, all magnetized planets in our solar system produce coherent radio emission by auroral processes powered by electron cyclotron maser instability, typically through magnetic reconnection between the planetary field and the field carried by the solar wind, as is the case with Earth, Saturn, Uranus, and Neptune. Jupiter's primary radio emission is driven by the solar wind compressing its co-rotating plasma sheath and by interactions with the magnetic fields of its moons. In all cases, the characteristic frequency of radio emission is proportional to the electron cyclotron frequency, which is determined by the magnetic field strength: $\nu_e \cong 2.8\, B_{Gauss}$ MHz. Jupiter's strong ~14 Gauss magnetic field produces radio emission with peak frequency of ~40 MHz. All other magnetized planets in our solar system have magnetic fields < 2 Gauss, making their peak emission frequencies below 6 MHz. Earth's aurora kilometric radiation (AKR; see Figure 1) peaks around ~0.3 MHz.

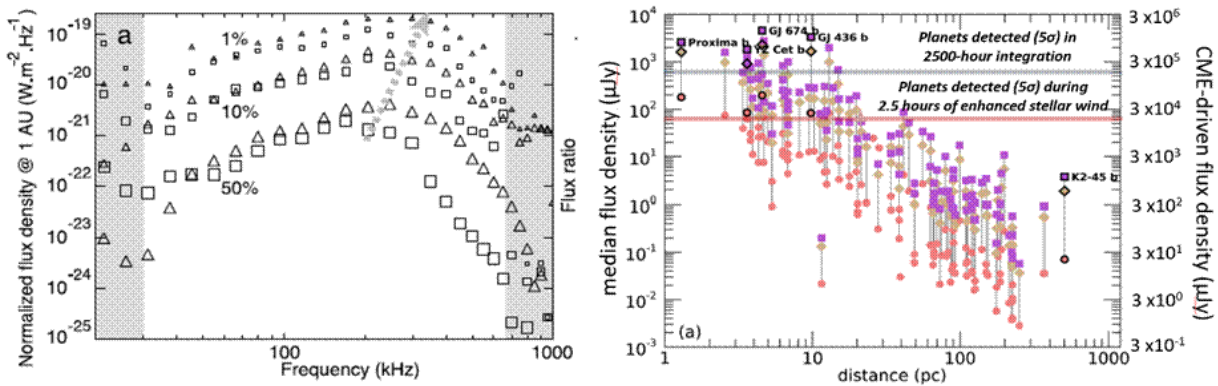

*Figure 1. (left) Radio spectrum of Earth's aurora kilometric radiation (AKR) for top 50%, 10%, and 1% of measurements. Flux density increases by nearly 1000x due to enhanced solar winds. (right) Predicted flux density of known exoplanets in M-dwarf systems (Vidotto et al. 2019) assuming Earth-like magnetic field of 0.1 Gauss. Colors reflect different models of the quiescent stellar wind (Burns et al. 2021).*

Detecting terrestrial-like exoplanet radio emission requires a large collecting area on a space-based observatory to avoid Earth's opaque ionosphere (below 10 MHz) and AKR. **The non-polar lunar farside is the only suitable location in the inner solar system since it is shielded from Earth's emissions** (Bassett et al. 2020). Lunar radio interferometric arrays, such as the FARSIDE pathfinder and larger Farview (Burns et al. 2021, Polidan et al. 2024), are under concept development to address this and other high-priority Astro2020 decadal science objectives. These lunar radio telescopes will survey large areas of the sky simultaneously to detect radio bursts by exoplanets in the solar neighborhood. FARSIDE and Farview are designed to image up to 10,000 square degrees simultaneously. These large fields will include 2000 stellar systems within 25 parsecs. For a telescope the size of FARSIDE, enhanced exoplanet radio emission during stellar CMEs will be detectable in a 2.5 hour observation, and the average (non-CME enhanced) radio emission from exoplanets within 10 parsecs will be measured over a nominal 2-year observing period. Stellar winds and mass ejections needed to feed the auroral activity can be sporadic, necessitating large fields of view and continuous surveying. Exoplanet emission will be identifiable by its circular polarization, distinguishable from bursts of the host star by rotational modulation, and not limited by classical source confusion because the interstellar medium plasma is optically thick at very low frequencies.

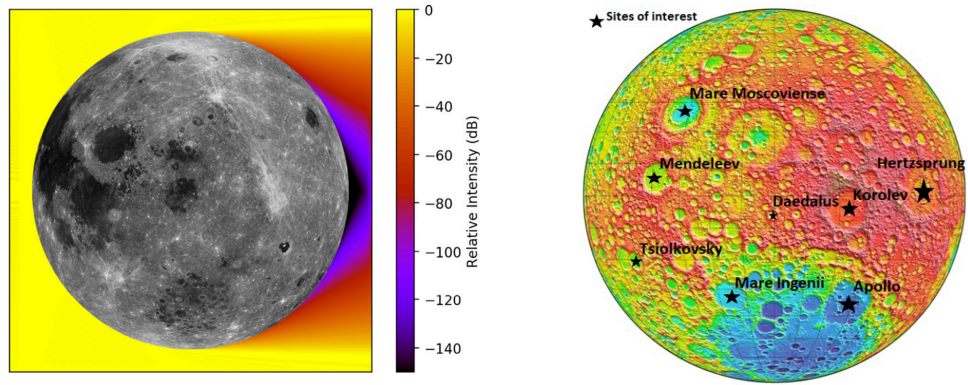

*Figure 2. (left) Electromagnetic model showing the suppression of AKR and Earth transmitters on the far side of the Moon (Bassett et al. 2020). (Right) Eight candidate sites on the farside identified in Krolikowski & Elvis 2024 for a lunar radio array considering available area and surface smoothness. Further work is necessary to downselect to a single site, potentially through contributions of human exploration to site surveys.*

**2. Human Exploration and Lunar Radio Telescope Sites:** Table 1 lists candidate non-polar lunar farside sites for radio telescopes. Prior to construction of the lunar radio telescopes, **human exploration is needed to provide detailed surveys** and characterization of the sites, which require smooth, flat surfaces over tens of kilometers with a low density of boulders and obstructions. The dielectric properties (permittivity and conductivity) of the regolith and subsurface layers to depths of several kilometers will need to be mapped with **ground penetrating radar (GPR)** to provide accurate electromagnetic modeling and polarization properties of the telescope antennas, and thermal characterization of the regolith will be needed to ensure strategies for burying electronics will be sufficient to enable continuous operations over many years. After construction, recurring human presence on the radio telescope site(s) will be critical to provide **maintenance and upgrades**—just as astronaut activities in LEO vastly extended the operational lifetime and performance of the Hubble Space Telescope.

| Table 1 – Lunar Farside Sites for Radio Telescopes | | |
|---|---|---|
| Site Name | Lunar Coordinate | Size (km) |
| Mare Moscoviense | 27.3°N, 147.9°E | 276 |
| Mendeleev | 5.7°N, 140.9°E | 313 |
| Hertzsprung | 1.37°N, 128.66°W | 570 |
| Korolev | 4.0°S, 157.4°W | 437 |
| Daedalus | 5.9°S, 179.4°E | 93 |
| Tsiolkovsky | 20.38°S, 128.97°E | 184 |
| Mare Ingenii | 33.7°S, 163.5°E | 282 |
| Apollo | 36.1°S, 151.8°W | 537 |

**3. References:** Bassett et al., Advances in Space Research, Volume 66, Issue 6, p. 1265-1275, 2020. Burns, J. O., et al., The Planetary Science Journal, Volume 2, 2, p. 44, 2021. Jakosky, B. M., et al., Science, vol. 350, no. 6261, p. aad0210, 2015. Hallinan, G., et al., Bulletin of the American Astronomical Society, Volume 53, 4, p. 379, 2021. Khodachenko, M. L., et al., Astrobiology, vol. 7, no. 1, pp. 167-184, 2007. Lammer, H. et al., Astrobiology, vol. 7, no. 1, pp. 185-207, 2007. Polidan, R. S., et al., Advances in Space Research, Volume 74, 1, p. 528, 2024. Vidotto, A. A., et al., Monthly Notices of the Royal Astronomical Society, Volume 488, 1, p. 633, 2019. Krolikowski. A, Elvis. M Phil. Trans. R. Soc. A. 2024